\documentclass[runningheads]{llncs}

\usepackage{ulem}
\usepackage{times}
\usepackage{subfig}
\usepackage{tabularx}
\usepackage{booktabs}
\usepackage{multirow}
\usepackage{xspace}
\newcolumntype{R}[1]{>{\raggedleft\arraybackslash\hspace{0pt}}p{#1}}
\newcolumntype{L}[1]{>{\raggedright\arraybackslash\hspace{0pt}}p{#1}}
\newcolumntype{C}[1]{>{\centering\hspace{0pt}}p{#1}}
\usepackage{tikz}
\usetikzlibrary{spy}
\usepackage{pgfplots}
\newlength \figureheight 
\newlength \figurewidth
\newcommand{\eg}{e.\,g.\xspace}
\newcommand{\etal}{\textit{et\,al.}\xspace}
\usepackage{array}

\usepackage{graphicx}
\usepackage{comment}
\usepackage{amsmath,amssymb}
\usepackage{color}
\usepackage{url}
\usepackage{hyperref}

\newif\ifreview
\reviewfalse

\ifreview
	\usepackage{lineno}

	\linenumbers
\fi

\begin{document}

\def\SubNumber{99}

\def\GCPRTrack{Regular Track}

\title{Merging-ISP: Multi-Exposure High Dynamic Range Image Signal Processing}

\ifreview
\else

	\author{
	Prashant Chaudhari\inst{1} \and
	Franziska Schirrmacher\inst{2} \and
	Andreas Maier\inst{3} \and
	Christian Riess\inst{2} \and
	Thomas Köhler\inst{1}
	}
	
	\authorrunning{P. Chaudhari \etal}
	
	\institute{
	e.solutions GmbH, Erlangen, Germany \and
	IT Security Infrastructures Lab, 
	Friedrich-Alexander-Universität (FAU) Erlangen-Nürnberg, Germany \and Pattern Recognition Lab, Friedrich-Alexander-Universität (FAU) Erlangen-Nürnberg, Germany \\
	\email{
	\{prashant.chaudhari,thomas.koehler\}@esolutions.de \\ \{franziska.schirrmacher,andreas.maier,christian.riess\}@fau.de
	}
	}
\fi

\maketitle              

\begin{abstract}
High dynamic range (HDR) imaging combines multiple images with different exposure times into a single high-quality image.
The image signal processing pipeline (ISP) is a core component in digital cameras to perform these operations. It includes demosaicing of raw color filter array (CFA) data at different exposure times, alignment of the exposures, conversion to HDR domain, and exposure merging into an HDR image. Traditionally, such pipelines cascade algorithms that address these individual subtasks. However, cascaded designs suffer from error propagation, since simply combining multiple steps is not necessarily optimal for the entire imaging task.

This paper proposes a multi-exposure HDR image signal processing pipeline (Merging-ISP) to jointly solve all these subtasks. Our pipeline is modeled by a deep neural network architecture. As such, it is end-to-end trainable, circumvents the use of hand-crafted and potentially complex algorithms, and mitigates error propagation. Merging-ISP enables direct reconstructions of HDR images of dynamic scenes from multiple raw CFA images with different exposures. We compare Merging-ISP against several state-of-the-art cascaded pipelines. The proposed method provides HDR reconstructions of high perceptual quality and it quantitatively outperforms competing ISPs by more than 1\,dB in terms of PSNR.
\end{abstract}

\section{Introduction} \label{sec:intro}

\begin{figure}[t]
    \centering
	\includegraphics[width=1.00\columnwidth]{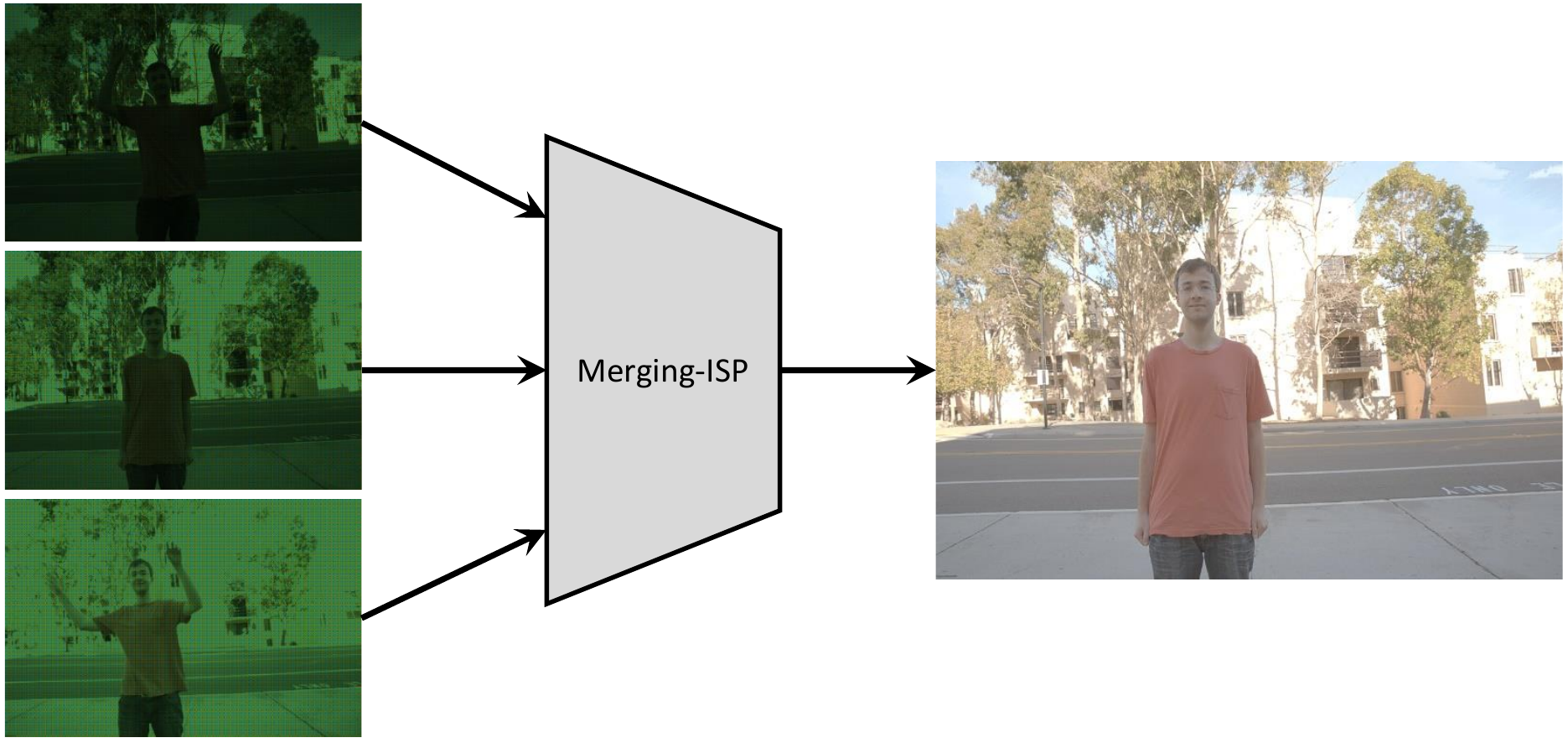}
	\caption{We propose a camera image signal processing pipeline (ISP) using deep neural networks to directly merge multi-exposure Bayer color filter array data of low dynamic range (LDR) into a high dynamic range (HDR) image.}
	\label{fig:teaser:2}
\end{figure}

Computational photography aims at computing displayable images of high perceptual quality from raw sensor data.
In digital cameras, these operations are performed by the image signal processing pipeline (ISP). The ISP plays a particularly important role in commodity devices like smartphones, to overcome physical limitations of the sensors and the optical system. ISPs cascade processing steps that perform different computer vision tasks. Initial steps address the low-level \textit{reconstruction} of images in RGB space from raw data, comprising demosaicing, denoising, and deblurring. Specifically, demosaicing \cite{Li2008} creates full RGB images from sensor readouts behind a color filter array (CFA) that provides per pixel only a single spectral component. Later steps in the ISP comprise higher level \textit{enhancement} operations like high dynamic range (HDR) imaging \cite{Kalantari2017,Sen2012,Wu2018}. One effective approach adopted in this paper enhances dynamic ranges by merging multiple low dynamic range (LDR) images, where each LDR image is captured sequentially with different exposure time~\cite{tomaszewska2007image}. These exposures have additionally to be aligned in an intermediary step when the scene is dynamic.

For each of these individual processing tasks, there exist several algorithms with different quality criteria and computational requirements. This enables camera manufacturers to assemble an ISP that is tailored to the capabilities of the device. Nevertheless, the resulting ISP is not necessarily optimal. In particular, error propagation between these isolated processing steps oftentimes leads to suboptimal results~\cite{Heide2014}. For example, demosaicing artifacts can be amplified by image sharpening, or misalignments of different exposures can reduce the overall image quality. 

Solving all tasks \textit{jointly} is a principal approach to overcome error propagation. This can be achieved by formulating the ISP output as solution of an inverse problem \cite{Heide2014}. The individual steps in the pipeline are modeled by a joint operator and signal processing is accomplished via non-linear optimization. While this yields globally optimal solutions, \eg in a least-squares sense, analytical modeling for real cameras is overly complex. Also uncertainties due to simplifications cause similar effects as error propagation. 

\textit{End-to-end learning} of the ISP \cite{Liang2019,Ratnasingam2019,Schwartz2018} avoids the need for analytical modeling. However, current methods in that category consider single exposures only. Different to merging multi-exposure data, they can only hallucinate HDR content. For instance, single-exposure methods often fail in highly over-saturated image regions \cite{Wu2018}. Aligning exposures captured sequentially from dynamic scenes is, however, challenging due to inevitable occlusions and varying image brightness. This is particularly the case when using raw CFA data preventing the use of standard image alignment.

\subsubsection{Contributions.}
This paper proposes a \textit{multi-exposure  high  dynamic range image  signal  processing  pipeline (Merging-ISP)} using a deep neural network architecture and end-to-end learning of its processing steps. Our method directly maps raw CFA data captured sequentially with multiple exposure times to a single HDR image, see Fig.~\ref{fig:teaser:2}. Contrary to related works \cite{Schwartz2018,Wu2018}, it jointly learns the alignment of multi-exposure data in case of dynamic scenes along with low-level and high-level processing tasks. As such, our Merging-ISP avoids common error propagation effects like demosaicing artifacts, color distortions, or image alignment errors amplified by HDR merging.

\section{Related Work} \label{sec:related}

Overall, we group ISPs into two different modules: 1) low-level vision aiming at image reconstruction on a pixel-level of CFA data to form full RGB images, and 2) high-level vision focusing on the recovery of HDR content from LDR observations. Our method also follows this modular design but couples all stages by end-to-end learning. 

\subsection{Low-Level Vision}

Low-level vision comprises image reconstruction tasks like demosaicing of CFA data, defect pixel interpolation, denoising, deblurring, or super-resolution. Classical pipelines employ isotropic filters (\eg, linear interpolation of missing CFA pixels or linear smoothing). Edge-adaptive techniques \cite{Menon2007,Tsai2007} can avoid blurring or zippering artifacts of such non-adaptive filtering. Another branch of research approaches image reconstruction from the perspective of regularized inverse problems with suitable image priors \cite{Koehler2016,Pan2016}. These methods are, however, based on hand-crafted models. Also simply cascading them leads to accumulated errors.

Later, deep learning advanced the state-of-the-art in denoising \cite{Zhang2017}, demosaicing \cite{Tan2018}, deblurring \cite{Tao2018}, or super-resolution \cite{Lai2018}. Deep neural networks enable image reconstructions under non-linear models. Using generative adversarial networks (GANs) \cite{Kupyn2018,Wang2018} or loss functions based on deep features \cite{Waleed2018} also allow to optimize such methods with regard to perceptual image quality. It also forms the base for multi-task learning of pipelines like demosaicing coupled with denoising \cite{Dong2018,Kokkinos2018} or super-resolution \cite{Zhou2018}. In contrast to cascading these steps, this can avoid error propagation. We extend this design principle by incorporating high-level vision, namely HDR reconstruction.

\subsection{High-Level Vision}

High-level vision focuses on global operations like contrast enhancement. In this context, we are interested in capturing HDR data. This can be done using special imaging technologies, \eg beam splitter \cite{Tocci2011} or coded images \cite{Serrano2016}, but such techniques are not readily available for consumer cameras due to cost or size constraints. There are also various inverse tone mapping methods estimating HDR data from single LDR acquisitions \cite{Eilertsen2017,Endo2017,Lee2018,Lee2018a}. However, this can only hallucinate the desired HDR content.

The dynamic range can also be increased by aligning and merging frame bursts. Merging bursts captured with constant exposure times \cite{Hasinoff2016,Zhang2010} enhances the signal-to-noise ratio and allows to increase dynamic ranges to a certain extent, while making alignment for dynamic scenes robust. In contrast, bursts with bracketed exposure images as utilized in this paper can expand dynamic ranges by much larger factors but complicates alignment. Therefore, multi-exposure alignment has been proposed \cite{Gallo2015,Jacobs2008,tomaszewska2007image}, which is especially challenging in case of large object motions or occlusions.

Recently, joint alignment and exposure merging have been studied to improve robustness. Patch based approaches ~\cite{Hu2013,Sen2012} fill missing under/over-exposed pixels in a reference exposure image using other exposures. Recent advances in deep learning enable merging with suppression of ghosting artifacts. In \cite{Kalantari2017}, Kalantari and Ramamoorthi have proposed a CNN for robust merging. Wu \etal \cite{Wu2018} have proposed to tackle multi-exposure HDR imaging as image translation. Their U-Net considers alignment as part of its learning process. However, existing patch-based or learning-based methods necessitate full RGB inputs and cannot handle raw CFA data directly. With Merging-ISP, we aim at direct HDR reconstruction from multiple raw images.

\subsection{End-to-End Coupling of Low-Level and High-Level Vision}

Several attempts have been made to couple low-level and high-level vision in an end-to-end manner. FlexISP \cite{Heide2014} is a popular model-based approach handling different sensors with their pixel layouts, noise models, processing tasks, and priors on natural images in a unified optimization framework. However, FlexISP and related approaches require analytical modeling of a given system as inverse problem, which can become complex. 

Data-driven approaches learn an ISP from example data and circumvent this effort. The DeepISP as proposed by Schwartz \etal \cite{Schwartz2018} enables direct mappings from low-light raw data to color corrected and contrast enhanced images. Ratnasingam \cite{Ratnasingam2019} has combined defect pixel interpolation, demosaicing, denoising, white balancing, exposure correction, and contrast enhancement using a single CNN. In contrast to monolithic networks, CameraNet introduced by Liang \etal \cite{Liang2019} comprises separate modules for these low-level reconstruction and high-level enhancement tasks. These frameworks are closely related to our proposed method but did not consider the reconstruction of HDR content. Our Merging-ISP uses multi-exposure data captured in burst mode for true HDR reconstruction rather than hallucinating such content from single images.

\section{Proposed Merging-ISP Framework} \label{sec:method}

\begin{figure}[!t]
	\centering
	\includegraphics[width=1.00\columnwidth]{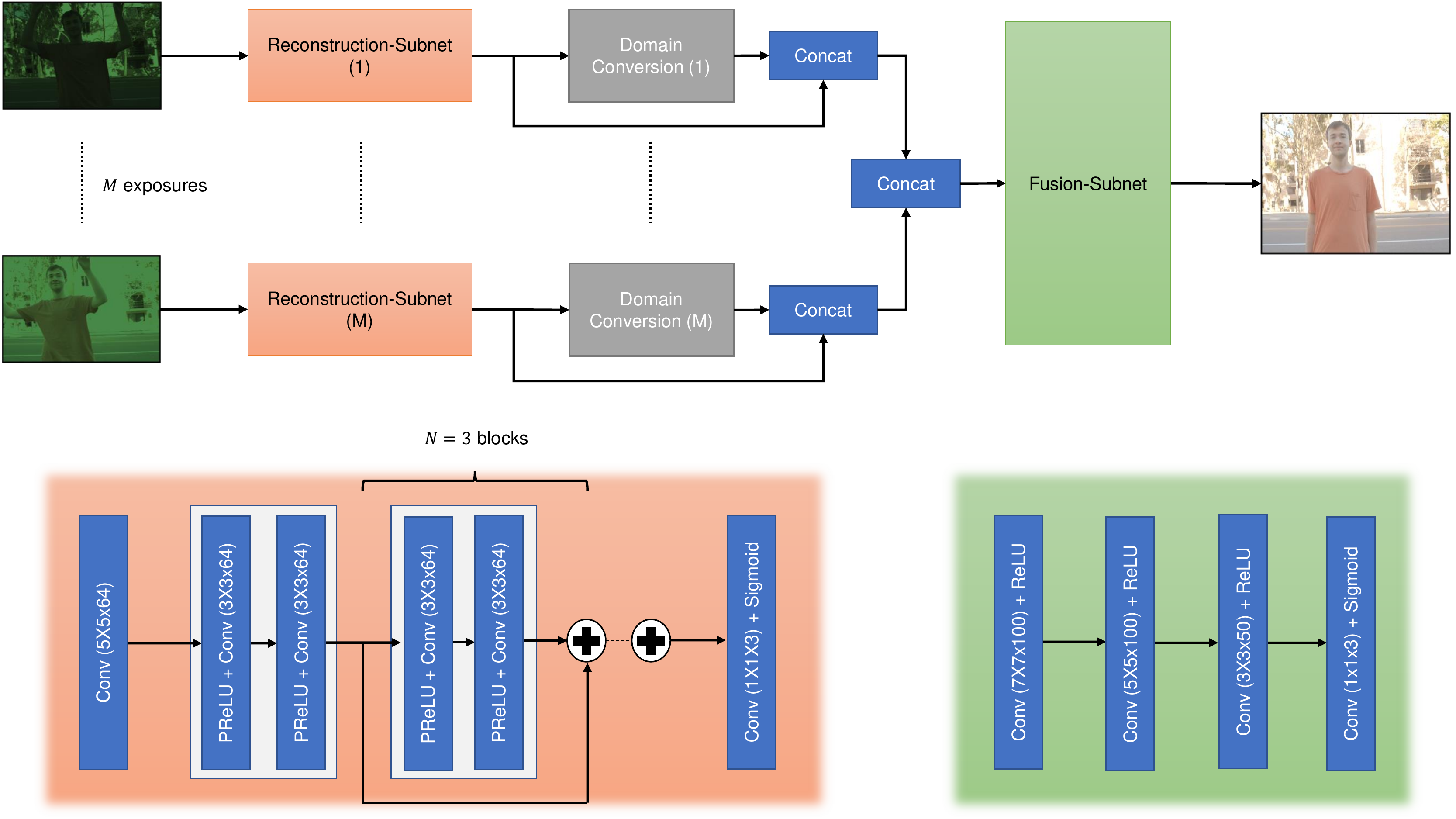}%
	\caption{Our overall Merging-ISP architecture. \textit{Reconstruction-Subnets} (orange block) map raw CFA inputs into an intermediate feature space. \textit{Domain conversion} transfers intermediate LDR features to HDR domain. \textit{Fusion-Subnet} uses a series of convolutional layers to align and merge the channel-wise concatenated features from the previous stage into one HDR image.}
	\label{fig:architecture}
\end{figure}

In this section, we introduce our Merging-ISP. The input to our pipeline is a stack of $M$ raw images $\mathbf{Y}_i~\in~\mathbb{R}^{N_x \times N_y \times 3}$, $i = 1, \ldots, M$ captured with a known CFA (\eg, Bayer pattern) and ascending exposure times. The CFA masks pixels in RGB space according to its predefined pattern. Also, each raw image is defined in LDR domain with limited bit depth. We further consider the challenging situation of \textit{dynamic scenes}, where these raw images are misaligned due to camera and/or object motion during their acquisition. 

Given the raw inputs in LDR domain, we aim at learning the mapping:
\begin{equation} 
	\label{eq:HDRISPProblemForm}
	\mathbf{X} = f(\mathbf{Y}_1, \ldots, \mathbf{Y}_M)\enspace,
\end{equation}
where $\mathbf{X} \in \mathbb{R}^{N_x \times N_y \times 3}$ is the target HDR image. To regularize the learning process of this general model, it is constrained such that the output HDR image is geometrically aligned to one of the inputs acting as a \textit{reference exposure}. For instance, we can choose the medium exposure to be the reference. Other exposures are aligned with this reference, which is done either separately or as integral part of the pipeline. In classical approaches, Eq.~\eqref{eq:HDRISPProblemForm} is further decomposed as $f(\mathbf{Y}) = (f_1 \odot f_2 \odot \ldots \odot f_K)(\mathbf{Y})$. This represents a camera ISP comprising $K$ successive processing stages $f_j(\mathbf{Y})$ that can be developed independently. In contrast, we learn the entire pipeline including the alignment of different exposures for dynamic scenes in an end-to-end manner.

\subsection{Network Architecture} \label{subsec:NetworkArchitecture}

Merging-ISP employs a modularized deep neural network to model the LDR-to-HDR mapping in Eq.~\eqref{eq:HDRISPProblemForm}. Figure \ref{fig:architecture} illustrates its architecture comprising three modules: parallel \textit{Reconstruction-Subnets} to restore intermediate features from input CFA data for each exposure, \textit{domain conversion} to transfer these features from LDR to HDR domain, and the \textit{Fusion-Subnet} estimating the final HDR output image from the feature space.

\subsubsection{Reconstruction-Subnet.}
This module implements the low-level vision tasks in our pipeline. Each raw CFA image is fed into one Reconstruction-Subnet instance, which consists of three fundamental stages as shown in Fig.~\ref{fig:architecture}. This subnet follows the general concept of a feed-forward
denoising convolutional neural network (DnCNN) \cite{DnCNN} and is a variation of the residual architecture in~\cite{Kokkinos2018}. The first stage is a single convolutional layer with 64 filters of size $5\times5$. The basic building block of the second stage are two layers with parametric rectified linear unit (PReLU) activation followed by 64 convolutional filters of size $3\times3$. We always use one of these blocks, followed by $N$ further blocks with skip connections. In our approach, we use $N=3$. The final stage comprises a single convolution layer with 3 filters of size $1\times1$ and sigmoid activation. The role of this final stage is to reduce the depth of the feature volume of the second stage from 64 channels to 3-channels $\mathbf{Z}_{i}$ for each exposure.
Note that \cite{Kokkinos2018} additionally links the input to that last layer, which we omit because we are fusing feature maps instead of restoring images.
We use reflective padding for all convolutions such that $\mathbf{Z}_{i}$ has the same spatial dimensions as the raw CFA input.

Feature volumes provided by the Reconstruction-Subnet encodes demosaiced versions of raw inputs in an intermediate feature space. Our Merging-ISP employs instances of this network to process the different exposures independently without parameter sharing. As a consequence of this design choice, the pipeline learns the reconstruction of full RGB data with consideration of exposure-specific data characteristics.

\subsubsection{Domain conversion.}
As Reconstruction-Subnet features are defined in LDR while exposure merging needs to consider HDR we integrate domain conversion as intermediate stage. Following the notion of \textit{precision learning} \cite{Maier2019}, we propose to formulate domain conversion using known operators. Since such existing operators are well understood, this can greatly reduce the overall number of trainable parameters and maximum error bounds of model training and thus the learning burden. Using the LDR feature volume $\mathbf{Z}_i$ obtained from the $i^{th}$ exposure $\mathbf{Y}_i$ and the conversion rule proposed in \cite{Kalantari2017}, the corresponding feature volume $\mathbf{Z}^H_i$ in the HDR domain is computed element-wise:
\begin{equation} 
\label{eq:LDRtoHDR}
Z^{H}_{i,jkl} = \frac{{Z_{i,jkl}}^\gamma}{t_i}\enspace,
\end{equation}
where $t_i$ is the exposure time and $\gamma = 2.2$. This domain conversion does not involve additional trainable parameters, which greatly simplifies the training of our Merging-ISP. It is applied to each of the intermediate feature volumes provided by the Reconstruction-Subnets.

\subsubsection{Fusion-Subnet.}
This is the high-level stage of our pipeline reconstructing HDR images. We construct its input by channel-wise concatenating LDR and HDR features as $\mathbf{U}_i = \mathrm{concat}(\mathbf{Z}_i, \mathbf{Z}^H_i)$ for each exposure. Given $M$ exposures and 3-channel images, we obtain the $N_x \times N_y \times 6M$ volume $\mathbf{U} = \mathrm{concat}(\mathbf{U}_1, \ldots, \mathbf{U}_M)$. This combined feature space facilitates the detection and removal of outliers like oversaturations to obtain artifact-free HDR content. Given the feature volume $\mathbf{U}$, the purpose of the Fusion-Subnet is a joint exposure alignment \textit{and} HDR merging, which is translated into two subtasks: 1) Its input feature volume $\mathbf{U}$ needs to be aligned towards one reference exposure coordinate frame to compensate for motion in dynamic scenes. 2) The aligned features need to be fused into the output HDR image $\mathbf{X}$. Both subtasks are solved intrinsically by the Fusion-Subnet.

We adopt the CNN proposed in \cite{Kalantari2017} for aligning and fusing multiple exposures in the Fusion-Subnet. Overall, the Fusion-Subnet comprises four convolutional layers with decreasing receptive fields of $7 \times 7$ for 100 filters in the first layer to $1 \times 1$ for 3 filters in the last layer. The input layer and hidden layers use rectified linear unit (ReLU) activation, while the output layer uses sigmoid activation to obtain linear-domain HDR data. Like in the Reconstruction-Subnet, each convolution uses reflective padding to preserve the spatial dimensions. One distinctive property compared to the baseline CNN in \cite{Kalantari2017} is that we do not employ an additional pre-alignment of input exposures in the image space, e.g., via optical flow. Instead we allow the Fusion-Subnet to learn this alignment in the feature space formed by the volume $\mathbf{U}$, which can greatly reduce error propagation in the entire ISP (see Section~\ref{sec:ablationStudy}).

\subsection{Loss Function} \label{subsec:LossFunctions}

While our proposed ISP provides predictions in a linear domain, HDR images are usually displayed after tonemapping. Hence, to train our model, we compute the loss on tonemapped HDR predictions. For tone mapping, we use the $\mu$-law \cite{Kalantari2017} defined as:
\begin{equation} 
\label{eq:tonemapping}
T(\mathbf{X}) = \frac{\log(1 + \mu\mathbf{X})}{\log(1 + \mu)}\enspace.
\end{equation}
Here, $\mu$ is a hyperparameter controlling the level of compression and $\mathbf{X}$ is an estimated HDR image in linear domain as inferred by Fusion-Subnet. We set $\mu = 5 \cdot 10^3$.

This differentiable tonemapping allows to formulate loss functions either on the basis of pixel-wise or perceptual measures \cite{Waleed2018}. In this paper, the overall loss function is defined pixel-wise using the $L_2$ norm:
\begin{equation} 
\label{eq:MSELossHDR}
\mathcal{L}(\tilde{\mathbf{X}}, \mathbf{X}) = \big|\big| T(\tilde{\mathbf{X}}) - T(\mathbf{X}) \big|\big|_2^2\enspace,
\end{equation}
where $\tilde{\mathbf{X}}$ denotes the ground truth and $T(\tilde{\mathbf{X}})$ is its tonemapped version.

\section{Experiments and Results}

In this section, we describe the training of the proposed Merging-ISP network and compare it against different ISPs for HDR imaging. In addition, we present an ablation study on the design of its deep neural network architecture.

\subsection{Datasets}

We include multiple databases to train and test our ISP ranging from synthetic to real data.

First, we use the HDR database collected by Kalantari \etal \cite{Kalantari2017} comprising ground truth HDR images with corresponding captured LDR images of 89 dynamic scenes. Each scene contains three bracketed exposure images and the medium exposure is used as a reference. Misalignments between these exposures are related to real non-rigid motion like head or hand movements of human subjects. To obtain raw CFA data as input for our ISP, we mosaic the LDR images with an RGGB mask~\cite{Khashabi2015}. Overall, 74 scenes are used for training or validation and the remaining 15 scenes are used for testing. 

To study the capability of our method to generalize to other sensors unseen during training, we additionally use examples of the multi-exposure dataset of Sen \etal~\cite{Sen2012}. The exposure values -1.3, 0, and +1.3 from each scene are used an input and the medium exposure is chosen as reference. The CFA for each exposure value employs an RGGB mask. This real-world dataset does not provide ground truths and is used for perceptual evaluation of HDR image quality in the wild.

\subsection{Training and Implementation Details} \label{sec:Experiments:Training}

Out of the 74 training scenes, we use 4 scenes to validate our model w.\,r.\,t.\xspace the loss function in Eq.~\eqref{eq:MSELossHDR}. Random flipping (left-right and up-down) and rotation by 90$^\circ$ of the images is performed to augment the training set from 70 to 350 scenes. Since training on full images has a high memory footprint, we extract 210,000 non-overlapping patches of size $50 \times 50$ pixels using a stride of $50$. The network weights are initialized using Xavier method \cite{Xavier2010} and training is done using Adam optimization \cite{Kingma2014} with $\beta_1 = 0.9$ and $\beta_2 = 0.999$. We perform training over 70 epochs with a constant learning rate of $0.0001$ and batches of size 32. During each epoch, all batches are randomly shuffled.

We implemented Merging-ISP in Tensorflow \cite{Abadi2015} with a NVIDIA GeForce 1080 Ti GPU. The overall training takes roughly 18 hours based on the architecture proposed in Section~\ref{subsec:NetworkArchitecture}. The prediction of one HDR image from three input CFA images with a resolution of $1500\times1000$ pixels takes 1.1 seconds. The performances of alternative network architectures considering modifications of the used residual blocks and convolutional layers are reported in our supplementary material.

\subsection{Comparisons with State-of-the-Art}

We compare our proposed Merging-ISP with several ISP variants that comprise different state-of-the-art demosaicing, denoising, image alignment, HDR merging, and tone mapping techniques. In terms of low-level vision, we evaluate directional filtering based demosaicing \cite{Menon2007} and deep joint demosaicing and denoising \cite{Kokkinos2018}. For the high-level stages, we use single-exposure HDR \cite{Eilertsen2017}, patch-based HDR \cite{Sen2012}, and learning-based HDR with a U-net \cite{Wu2018}. We use the publicly available source codes for all methods. 

We conduct our benchmark by calculating PSNR and SSIM of the ISP outputs against ground truth HDR data. In addition to pixel-based measures, we use HDR-VDP-2 \cite{Mantiuk2011} that expresses mean opinion scores based on the probability that a human observer notices differences between ground truths and predictions. HDR-VDP-2 is evaluated assuming 24 inches displays and 0.5\,m viewing distance. For fair qualitative comparisons, we show all outputs using Durand tonemapping with $\gamma = 2.2$ \cite{Durand2002}.

\begin{table}[!t]
	\centering
	\caption{Quantitative comparison of Merging-ISP to cascades of state-of-the-art demosaicing \cite{Menon2007,Kokkinos2018}, single-exposure HDR reconstruction \cite{Eilertsen2017} and multi-exposure HDR \cite{Sen2012,Wu2018} on the Kalantari test set. We report the mean PSNR, SSIM, and HDR-VDP-2 of output HDR images after tonemapping considering all combinations of cascaded demosaicing and HDR imaging.
	}
	\begin{tabular}{p{2.75cm}p{2.75cm}>{\centering\arraybackslash}p{1.75cm}>{\centering\arraybackslash}p{1.75cm}>{\centering\arraybackslash}p{1.75cm}}
		\toprule
		\textbf{Demosaicing} & \textbf{HDR Merging} & \textbf{PSNR} & \textbf{SSIM} & \textbf{HDR-VDP-2}  \\
		\midrule
		\multirow{3}[3]{*}{Menon \etal \newline \cite{Menon2007}} & Sen \etal \cite{Sen2012} & 28.37 & 0.9633 & 62.27 \\
		\cmidrule(l){2-5}
		& Eilertsen \etal \cite{Eilertsen2017} & 17.34 & 0.7673 & 53.36 \\
		\cmidrule(l){2-5}
		& Wu \etal \cite{Wu2018} & 28.28 & 0.9661 & 62.15 \\
		\midrule
		\multirow{3}{*}{Kokkinos \etal \cite{Kokkinos2018}} & Sen \etal \cite{Sen2012} & 40.07 & 0.9898 & 63.49 \\
		\cmidrule(l){2-5}
		& Eilertsen \etal \cite{Eilertsen2017} & 16.16 & 0.7689 & 53.81 \\
		\cmidrule(l){2-5}
		& Wu \etal \cite{Wu2018} & 41.62 & 0.9942 & 64.99 \\
		\midrule
		\multicolumn{2}{l}{Proposed Merging-ISP} & \textbf{43.17} & \textbf{0.9951} & \textbf{65.29} \\
		\bottomrule
	\end{tabular}
	\label{table:compareISP}
\end{table}

\begin{figure}[!t]
    \captionsetup[subfloat]{farskip=0.5pt,captionskip=1pt}
	\centering
	\subfloat[Raw inputs]{
		\includegraphics[width=0.25\columnwidth]{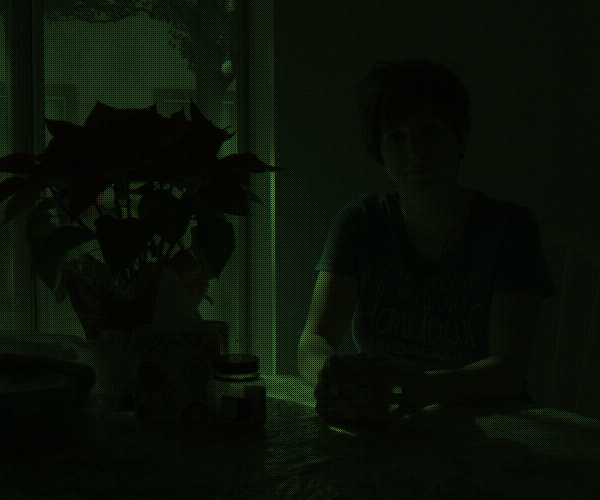}
		\includegraphics[width=0.25\columnwidth]{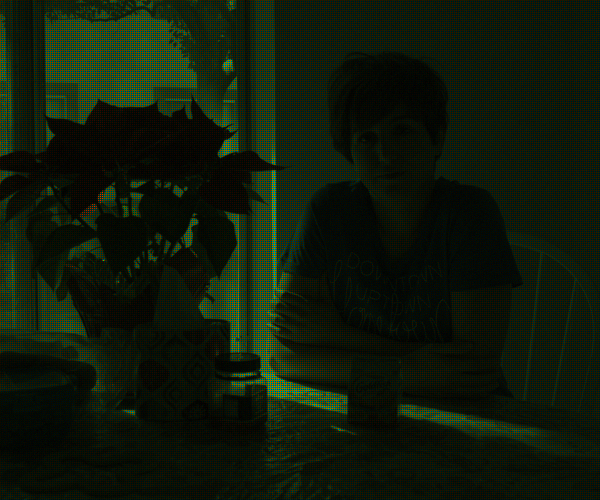}
		\includegraphics[width=0.25\columnwidth]{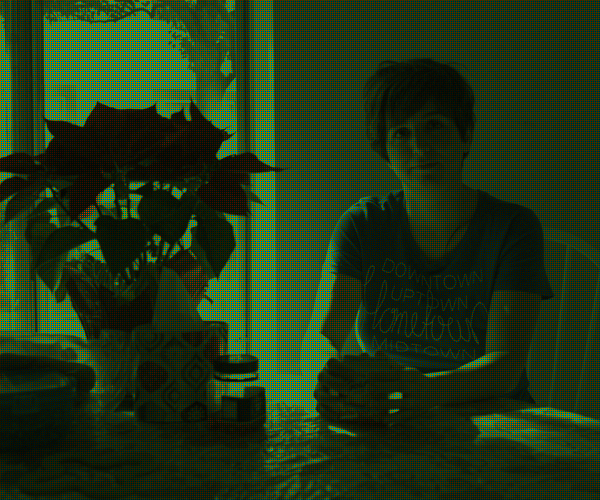}
	    \label{fig:ISPCompareKalantariSet:rawinputs}
	}
	
    \subfloat[Demosaicing \cite{Menon2007} + HDR \cite{Sen2012}]{
	    \begin{tikzpicture} [spy using outlines={rectangle,magnification=4, height=2.5cm, width=1.5cm, 
	        every spy on node/.append style={thick}}]
		    \node {\includegraphics[width=0.25\columnwidth]{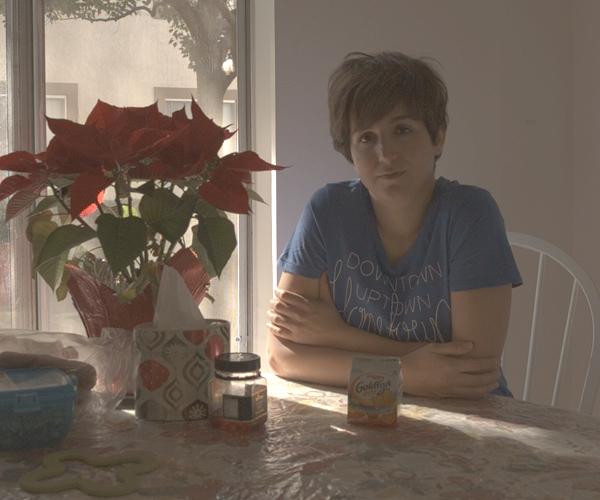}}; 
		    \spy[red] on (1, 0) in node [right] at (1.55, 0);
	    \end{tikzpicture}
    }
    \subfloat[Demosaicing \cite{Kokkinos2018} + HDR \cite{Wu2018}]{
		\begin{tikzpicture} [spy using outlines={rectangle,magnification=4, height=2.5cm, width=1.5cm,
			every spy on node/.append style={thick}}]
		\node {\includegraphics[width=0.25\columnwidth]{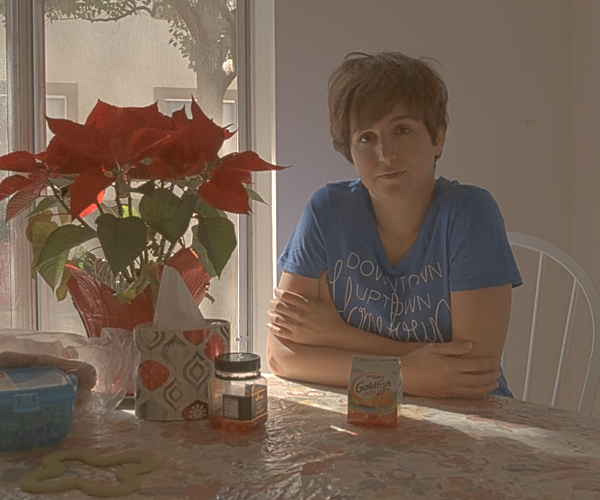}};
        \spy[red] on (1, 0) in node [right] at (1.55, 0);
	    \end{tikzpicture}
    \label{}
    }
    
    \subfloat[\centering Merging-ISP]{
		\begin{tikzpicture} [spy using outlines={rectangle,magnification=4, height=2.5cm, width=1.5cm,
			every spy on node/.append style={thick}}]
		\node {\includegraphics[width=0.25\columnwidth]{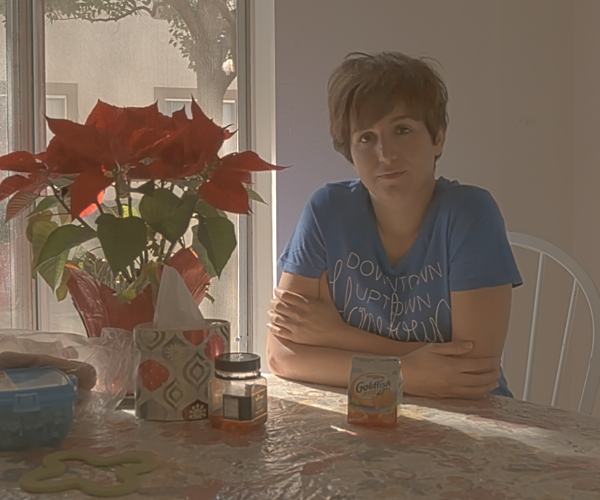}};
		\spy[red] on (1, 0) in node [right] at (1.55, 0);
	    \end{tikzpicture}
    }
    \subfloat[\centering Ground truth]{
		\begin{tikzpicture} [spy using outlines={rectangle,magnification=4, height=2.5cm, width=1.5cm,
			every spy on node/.append style={thick}}]
		\node {\includegraphics[width=0.25\columnwidth]{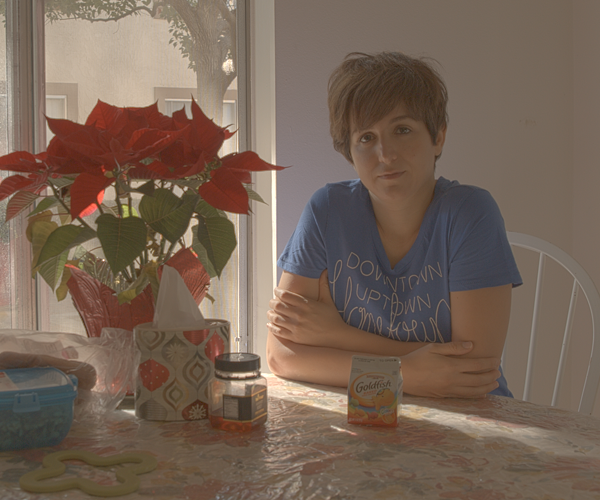}};
 		\spy[red] on (1, 0) in node [right] at (1.55, 0);
	    \end{tikzpicture}
    }
	\caption{
	Comparison of our Merging-ISP against different baseline ISPs formed by cascading state-of-the-art demosaicing \cite{Menon2007,Kokkinos2018} and multi-exposure HDR methods \cite{Sen2012,Wu2018}. The cascaded methods shown in (b) and (c) suffer from error propagations like demosaicing artifacts causing residual noise in the output. In contrast, Merging-ISP shown in (d) avoids noise amplifications.}
	\label{fig:ISPCompareKalantariSet}
\end{figure}

\begin{figure}[!t]
    \captionsetup[subfloat]{farskip=0.5pt,captionskip=1pt}
	\centering
	\subfloat[Raw inputs]{
		\includegraphics[width=0.25\columnwidth]{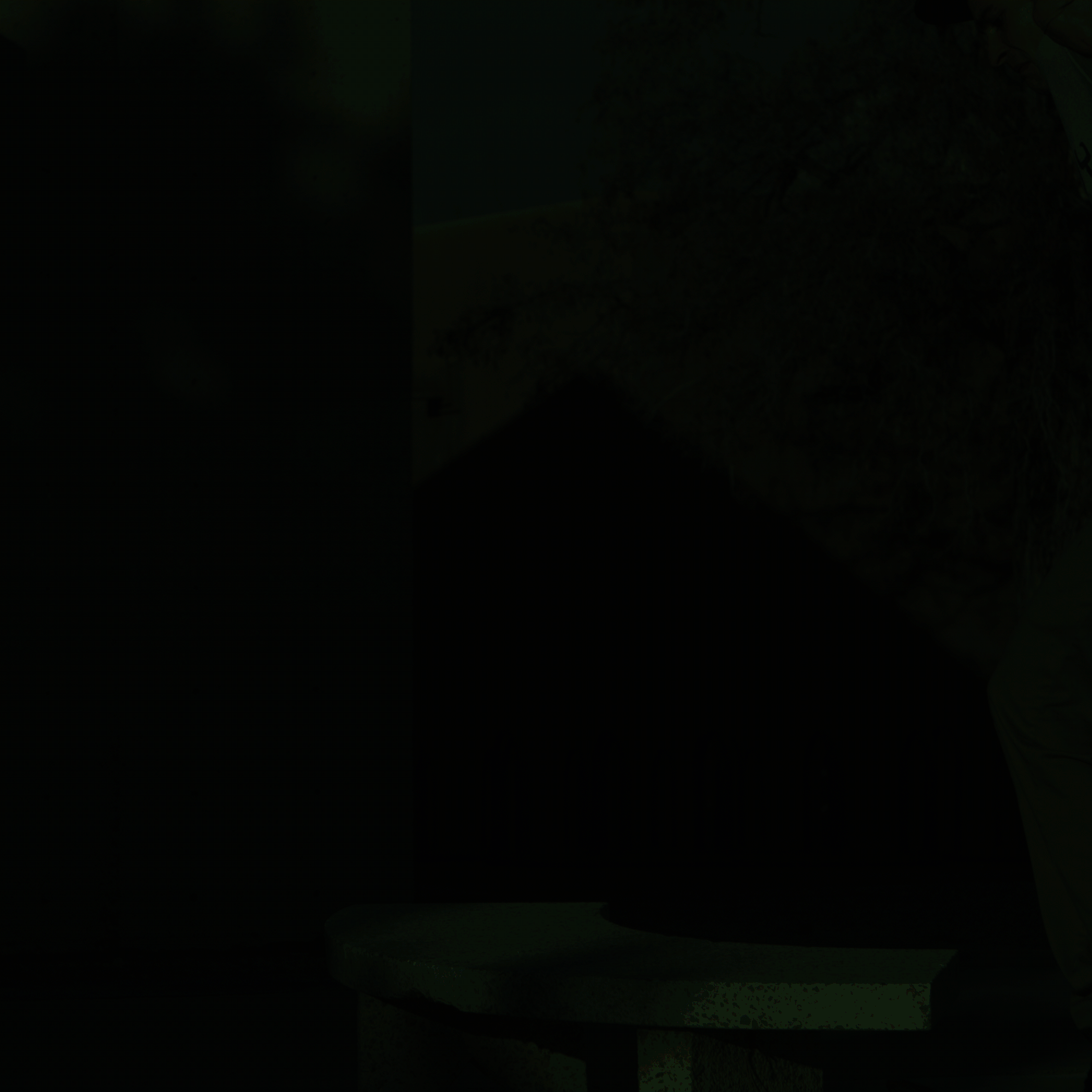}
		
		\includegraphics[width=0.25\columnwidth]{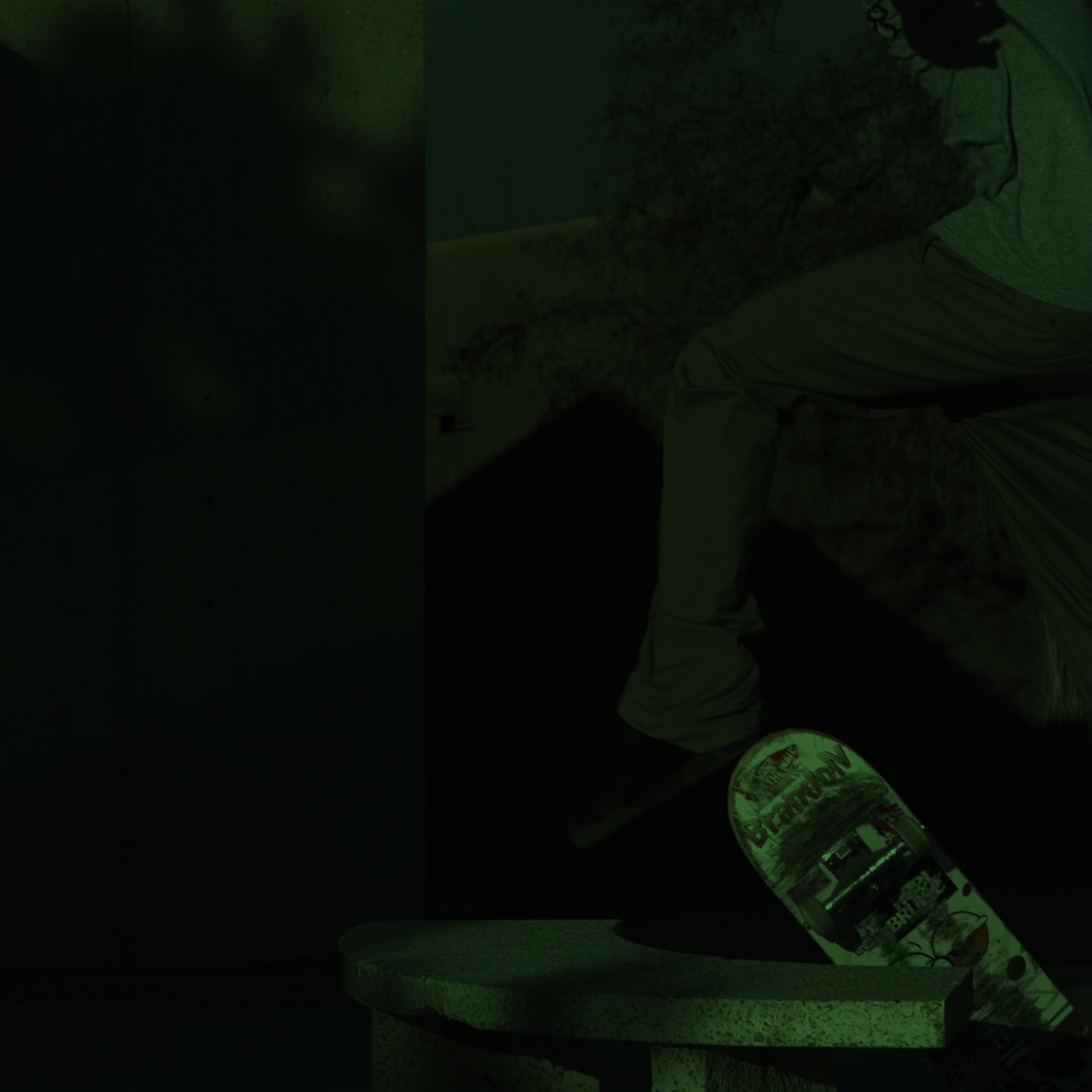}
		
		\includegraphics[width=0.25\columnwidth]{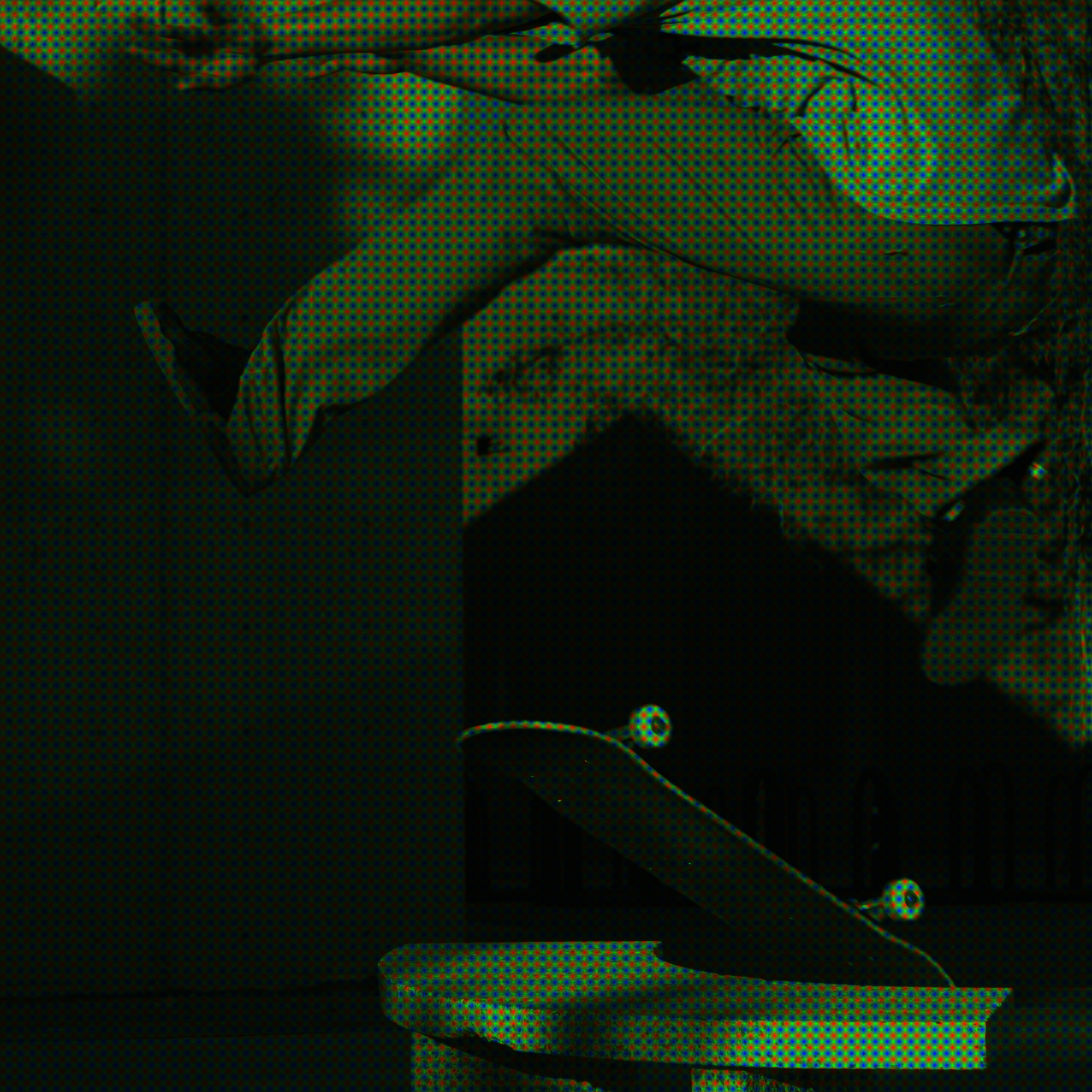}
	}
	
    \subfloat[\centering Demosaicing \cite{Kokkinos2018} + HDR \cite{Wu2018}]{
	    \begin{tikzpicture} [spy using outlines={rectangle,magnification=4.7, height=1.45cm, width=1.3cm, 
	        every spy on node/.append style={thick}}]
		    \node {\includegraphics[width=0.25\columnwidth]{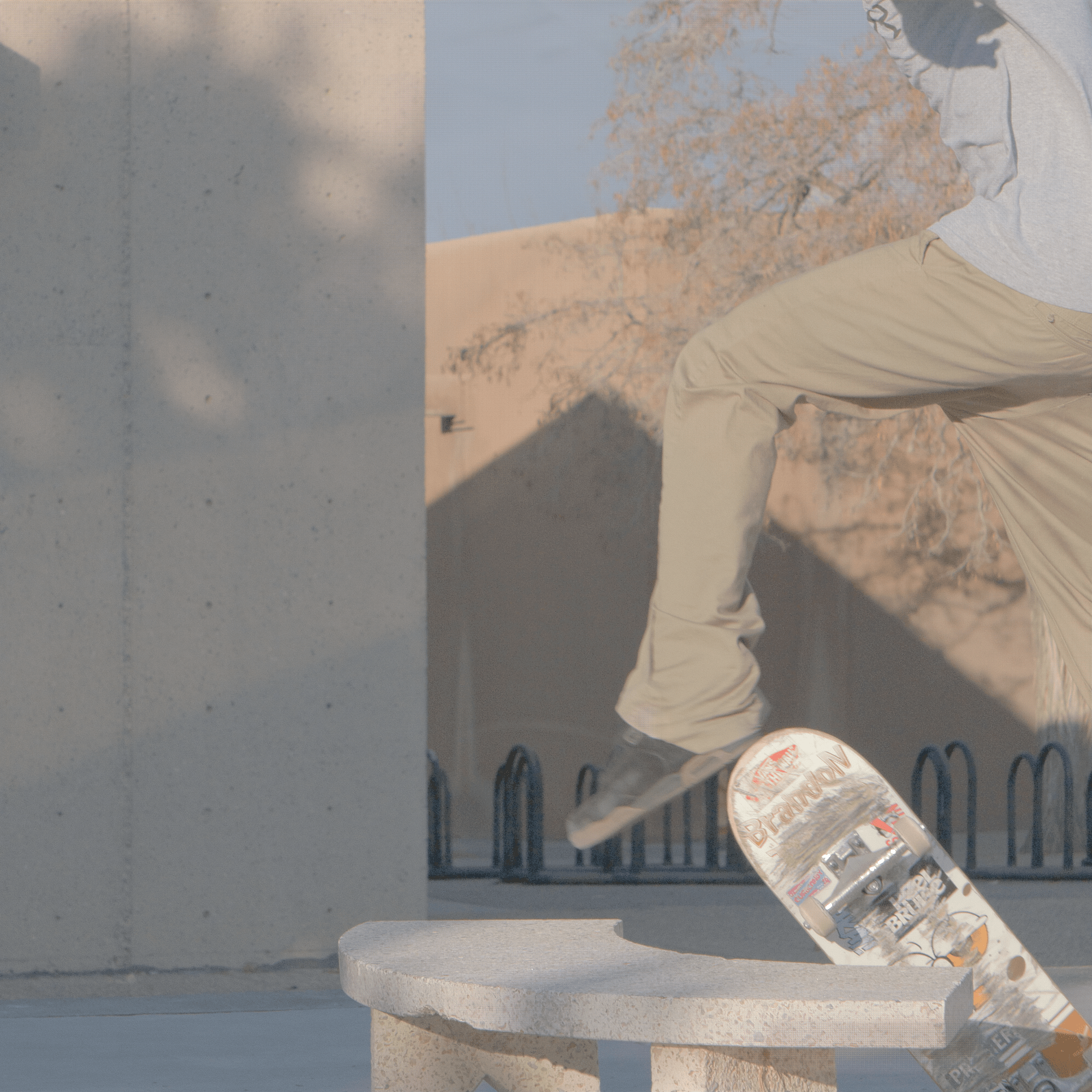}}; 
		    \spy[orange] on (-0.3, 1.2) in node [right] at (1.6, 0.8);
		    \spy[blue] on (-0.1, -0.75) in node [right] at (1.6, -0.8);
		    \spy[red] on (0.95, 1.35) in node [right] at (3, 0.8);
		    \spy[black] on (0.65, -0.7) in node [right] at (3, -0.8);
	    \end{tikzpicture}
    }
    \subfloat[Merging-ISP]{
	    \begin{tikzpicture} [spy using outlines={rectangle,magnification=4.7, height=1.45cm, width=1.3cm,
	        every spy on node/.append style={thick}}]
		    \node {\includegraphics[width=0.25\columnwidth]{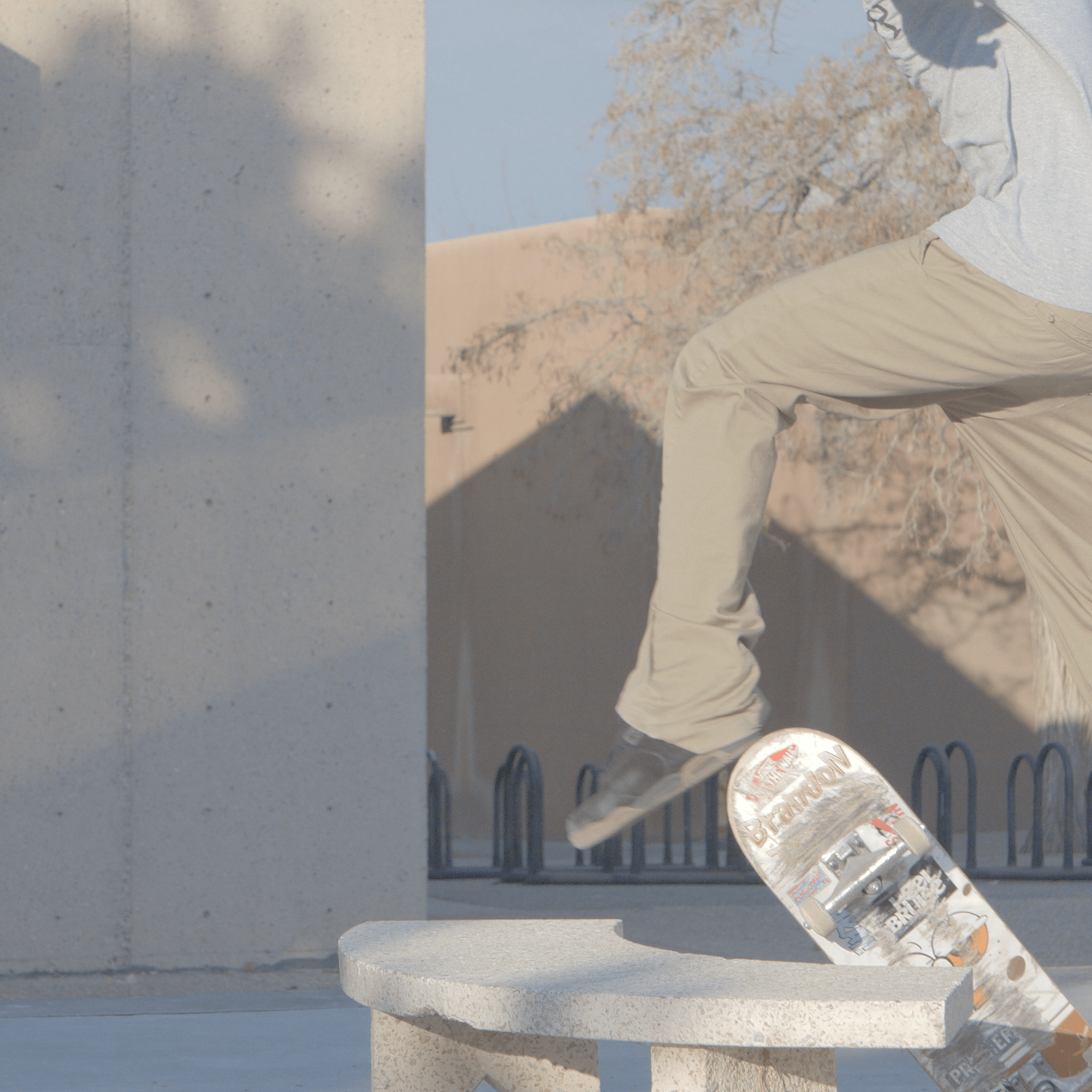}}; 
		    \spy[orange] on (-0.3, 1.2) in node [right] at (1.6, 0.8);
		    \spy[blue] on (-0.1, -0.75) in node [right] at (1.6, -0.8);
		    \spy[red] on (0.95, 1.35) in node [right] at (3, 0.8);
		    \spy[black] on (0.65, -0.7) in node [right] at (3, -0.8);
	    \end{tikzpicture}
    }
    \caption{Comparison of the best competing cascading ISP (demosaicing \cite{Kokkinos2018} + learning-based HDR \cite{Wu2018}) against Merging-ISP on a challenging scene from the dataset by Sen \etal~\cite{Sen2012}. The used Bayer pattern appears as artifact in the cascaded ISP output, which is mitigated by the Merging-ISP (see orange and red image patches).}
    \label{fig:SenDataset}
\end{figure}

\subsubsection{Comparison against cascaded ISPs.}

To the best of our knowledge, there are no related methods for direct reconstruction of HDR content from multiple CFA images. Thus, we compare against the design principle of cascading state-of-the-art algorithms for subtasks and combine different demosaicing/denoising methods with HDR merging.

In Table~\ref{table:compareISP}, we report mean quality measures of our method and six cascaded ISPs in the tonemapped domain on 15 scenes of the Kalantari test set. We found that Merging-ISP consistently outperforms the cascaded ISPs by a large margin. We compare on a test scene from the Kalantari dataset in Fig.~\ref{fig:ISPCompareKalantariSet}. Here, Merging-ISP provides HDR content of high perceptual quality, while the cascades suffer from error propagation. For example, even with the integration of state-of-the-art demosaicing/denoising like \cite{Kokkinos2018}, the cascaded designs lead to noise breakthroughs. 

We also compare Merging-ISP to the most competitive cascaded ISP variant (demosaicing \cite{Kokkinos2018} + HDR \cite{Wu2018}) on a challenging real raw scene from Sen \etal~\cite{Sen2012} in Fig.~\ref{fig:SenDataset}, where image regions are underexposed across all inputs and there is notable foreground motion between the exposures. Here, the cascaded design propagates demosaicing artifacts to exposure merging and thus to the final HDR output in static background (orange and red image patch). In the moving foreground (blue and black image patch), the cascaded design suffers from motion artifacts due to inaccurate alignments. Both types of error propagations are mitigated by Merging-ISP.
\newcolumntype{P}[1]{>{\arraybackslash}p{#1}}

\subsubsection{Comparison against single-exposure HDR imaging.}

\begin{figure}[ht!]
	\centering
	\makebox[0pt]{
		\subfloat{
			\begin{tikzpicture}
			\node {\includegraphics[width=0.175\columnwidth]{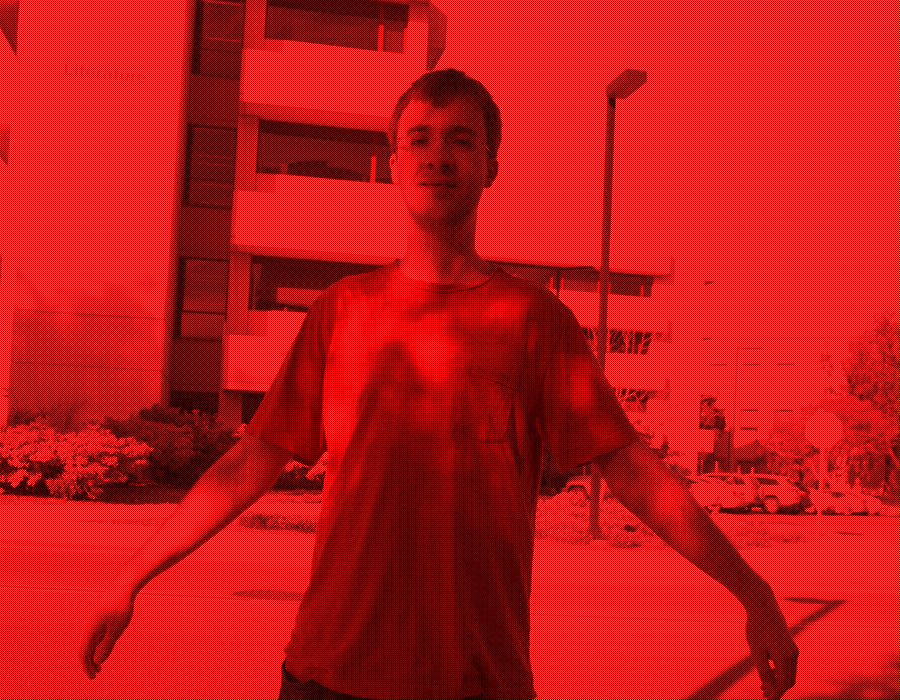}};
			\end{tikzpicture}
		}\hspace{-1.1em}
		\subfloat{
			\begin{tikzpicture} [spy using outlines={rectangle,magnification=4, height=1.65cm, width=1.2cm, every spy on node/.append style={thick}}]
			\node {\includegraphics[width=0.175\columnwidth]{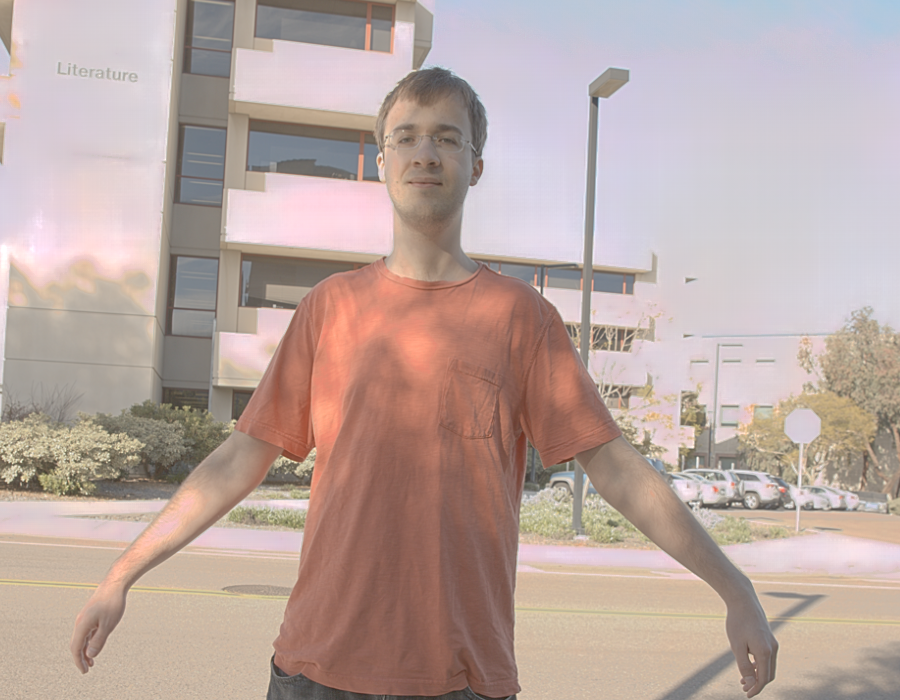}};
			\spy[blue] on (0.6, -0) in node [right] at (1.1, 0);
			\end{tikzpicture}
		}\hspace{-0.8em}
		\subfloat{
			\begin{tikzpicture} [spy using outlines={rectangle,magnification=4, height=1.65cm, width=1.2cm, every spy on node/.append style={thick}}]
			\node {\includegraphics[width=0.175\columnwidth]{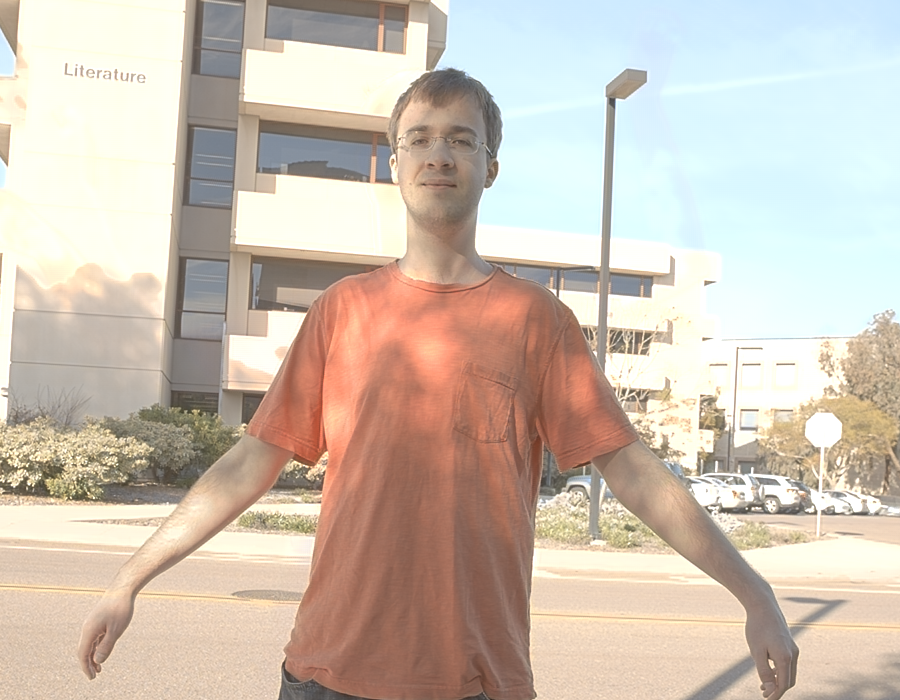}};
			\spy[blue] on (0.6, 0) in node [right] at (1.1, 0);
			\end{tikzpicture}
		}\hspace{-0.8em}
		\subfloat{
			\begin{tikzpicture} [spy using outlines={rectangle,magnification=4, height=1.65cm, width=1.2cm, every spy on node/.append style={thick}}]
			\node {\includegraphics[width=0.175\columnwidth]{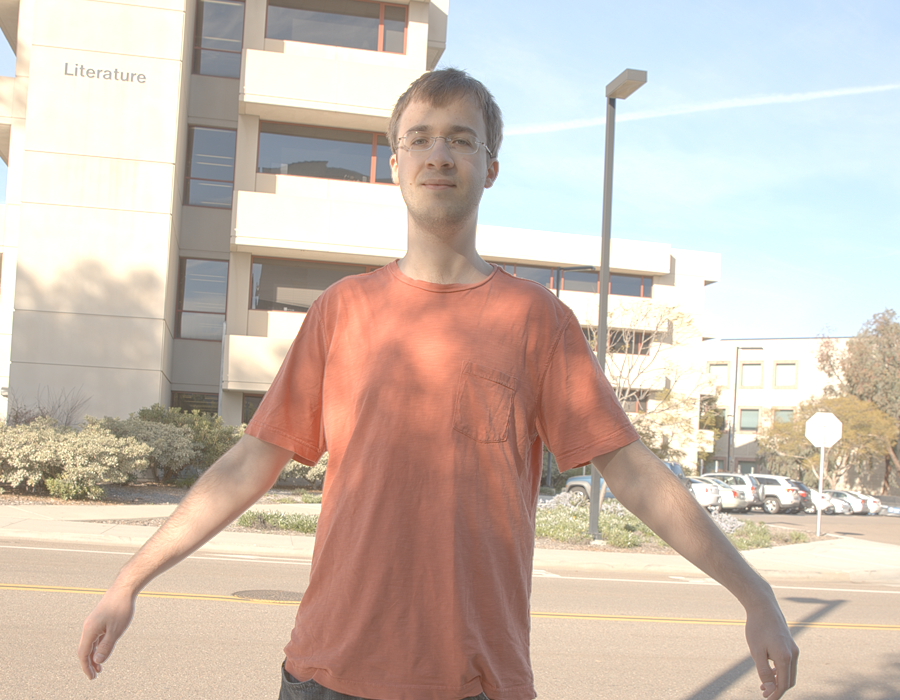}};
			\spy[blue] on (0.6, 0) in node [right] at (1.1, 0);
			\end{tikzpicture}
	}}\\[-0.5ex]
	\makebox[0pt]{
		\setcounter{subfigure}{0}
		\subfloat[Reference]{
			\begin{tikzpicture}
			\node {\includegraphics[width=0.175\columnwidth]{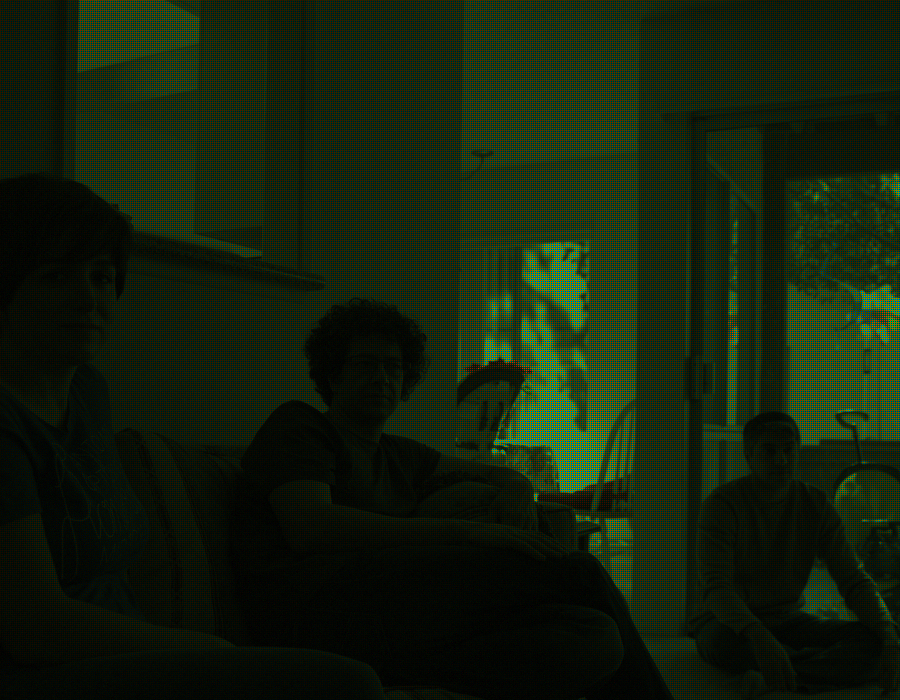}};
			\end{tikzpicture}
			\label{fig:SigleHDRCompare:INPUT}
		}\hspace{-1.1em}
		\subfloat[Demosaicing \cite{Kokkinos2018} + single-exposure HDR \cite{Eilertsen2017}][\centering Demosaicing \cite{Kokkinos2018} + \newline single-exposure HDR \cite{Eilertsen2017}]{
			\begin{tikzpicture} [spy using outlines={rectangle,magnification=4, height=1.65cm, width=1.2cm, every spy on node/.append style={thick}}]
			\node {\includegraphics[width=0.175\columnwidth]{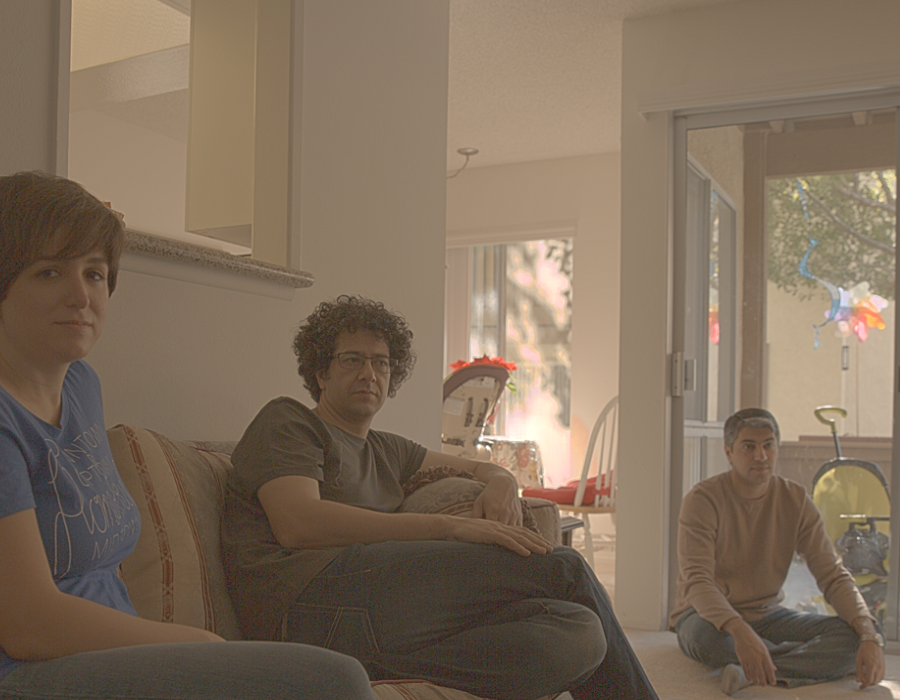}};
			\spy[blue] on (0.3, -0.3) in node [right] at (1.1, 0);
			\end{tikzpicture}
			\label{fig:SigleHDRCompare:HDRCNN}
		}\hspace{-0.8em}
		\subfloat[Merging-ISP]{
			\begin{tikzpicture} [spy using outlines={rectangle,magnification=4, height=1.65cm, width=1.2cm, every spy on node/.append style={thick}}]
			\node {\includegraphics[width=0.175\columnwidth]{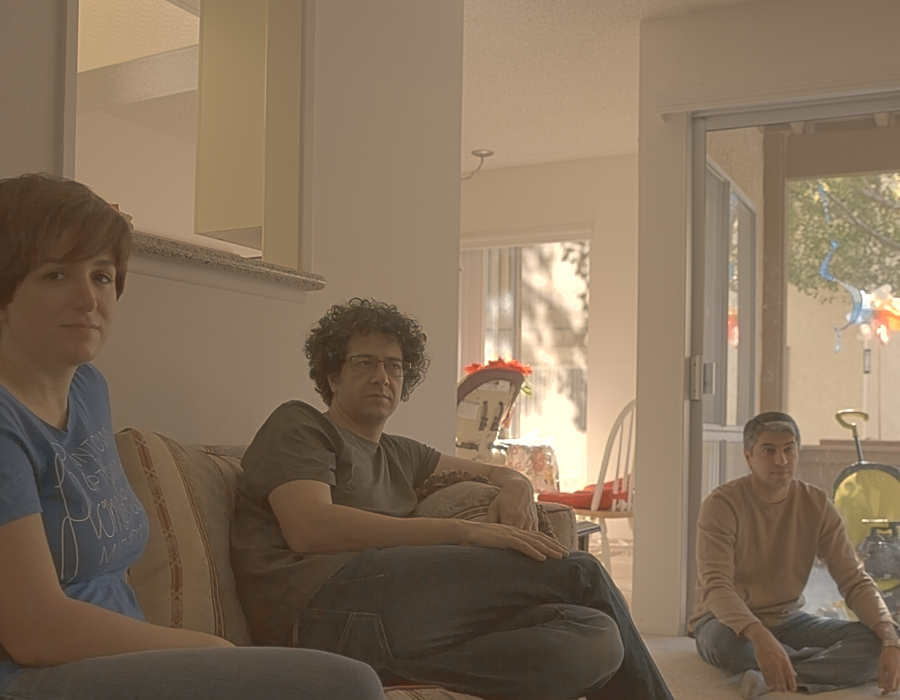}};
			\spy[blue] on (0.3, -0.3) in node [right] at (1.1, 0);
			\end{tikzpicture}
			\label{fig:SigleHDRCompare:HDRNET}
		}\hspace{-0.8em}
		\subfloat[Ground truth]{
			\begin{tikzpicture} [spy using outlines={rectangle,magnification=4, height=1.65cm, width=1.2cm, every spy on node/.append style={thick}}]
			\node {\includegraphics[width=0.175\columnwidth]{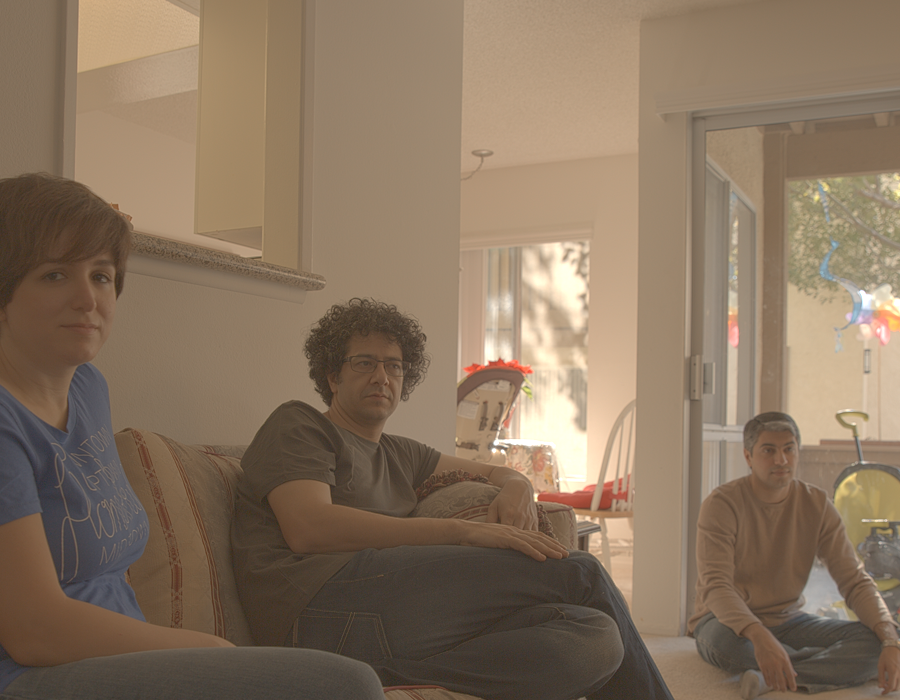}};
			\spy[blue] on (0.3, -0.3) in node [right] at (1.1, 0);
			\end{tikzpicture}
			\label{fig:SigleHDRCompare:GROUNDTRUTH}
	}}
	\caption{Comparison of our multi-exposure based Merging-ISP against state-of-the-art single-exposure HDR reconstruction \cite{Eilertsen2017} (cascaded with demosaicing \cite{Kokkinos2018}). In contrast to single-exposure methods, the proposed method mitigates color distortions.}
	\label{fig:SigleHDRCompare}
\end{figure}

In Fig.~\ref{fig:SigleHDRCompare}, we compare our multi-exposure approach against a recent deep learning method for single-exposure HDR reconstruction \cite{Eilertsen2017}. For fair comparisons, we demosaic the reference exposure raw image using the method in \cite{Kokkinos2018} and feed the preprocessed image into the HDR reconstruction developed in \cite{Eilertsen2017}. Overall, a single-exposure approach does not require alignments of multiple exposures in dynamic scenes. However, it fails to recover reliable color information, \eg in high saturated regions, as depicted in Fig.~\ref{fig:SigleHDRCompare:HDRCNN}. Merging-ISP in Fig.~\ref{fig:SigleHDRCompare:HDRNET} exploits multiple exposure inputs and avoids such color distortions.

\subsection{Ablation Study} 
\label{sec:ablationStudy}

We investigate the influence of different design choices of our Merging-ISP in an ablation study. To this end, we develop and compare several variants of this pipeline.

\begin{table}[t]
	\centering
	\footnotesize
	\caption{Ablation study of our Merging-ISP on the Kalantari data. We compare the proposed end-to-end learning of all subtasks against cascading our Reconstruction- and Fusion-Subnets with optical flow alignment \cite{Liu2009} before and after the reconstruction (pre- and post-align cascaded Merging-ISP) as well as end-to-end learning of both networks with optical flow pre-alignment (pre-align end-to-end Merging-ISP).}
	\begin{tabular}{l>{\centering\arraybackslash}p{1.75cm}>{\centering\arraybackslash}p{1.75cm}>{\centering\arraybackslash}p{1.75cm}}
		\toprule
		& \textbf{PSNR} & \textbf{SSIM} & \textbf{HDR-VDP-2} \\
		\midrule
		Pre-align cascaded Merging-ISP & 41.16 & 0.9923 & 64.71 \\
		\midrule
		Post-align cascaded Merging-ISP & 42.03 & 0.9937 & 65.09 \\
		\midrule
		Pre-align end-to-end Merging-ISP & 42.56 & 0.9945 & 64.84 \\
		\midrule
		Proposed Merging-ISP & \textbf{43.17} & \textbf{0.9951} & \textbf{65.29} \\
		\bottomrule
	\end{tabular}
	\label{table:AblationStudy}
\end{table}

\subsubsection{Learning subtasks separately vs. end-to-end learning.}
In our method, all ISP subtasks are learned in an end-to-end fashion.
To study the impact of this property, we evaluate variants of Merging-ISP, where subtasks are solved separately and cascaded to from an ISP. To this end, we trained the Reconstruction-Subnet for demosaicing and the Fusion-Subnet to merge demosaiced exposures. Different to our approach that intrinsically covers exposure alignment, the optical flow algorithm of Liu \cite{Liu2009} is used in two variants for alignment: optical flow can be either computed on CFA input data (referred to as \textit{pre-align cascaded} Merging-ISP) or on the Reconstruction-Subnet output (referred to as \textit{post-align cascaded} Merging-ISP). Both approaches can be considered as extensions of the HDR merging proposed in \cite{Kalantari2017} equipped with optical flow alignment and demosaicing implemented by our Reconstruction-Subnet to handle raw CFA inputs.

\begin{figure}[!t]
	\centering
	\subfloat{
		\includegraphics[width=0.19\linewidth]{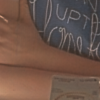}
	}
	\subfloat{
		\includegraphics[width=0.19\linewidth]{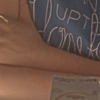}
	}
	\subfloat{
		\includegraphics[width=0.19\linewidth]{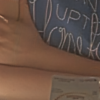}
	}
	\subfloat{
		\includegraphics[width=0.19\linewidth]{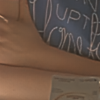}
	}
	\subfloat{
		\includegraphics[width=0.19\linewidth]{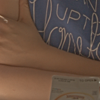}
	}\\
	\setcounter{subfigure}{0}
	\subfloat[Pre-align cascaded][\centering Pre-align \newline cascaded]{
		\includegraphics[width=0.19\linewidth]{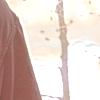}
		\label{fig:AblationStudy:PreAlignCascading}
	}
	\subfloat[Post-align cascaded][\centering Post-align \newline cascaded]{
		\includegraphics[width=0.19\linewidth]{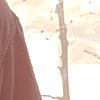}
		\label{fig:AblationStudy:PostAlignCascading}
	}
	\subfloat[Pre-align end-to-end][\centering Pre-align \newline end-to-end]{
		\includegraphics[width=0.19\linewidth]{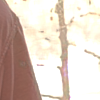}
		\label{fig:AblationStudy:PreAlignHDRNet}
	}
	\subfloat[Proposed]{
		\includegraphics[width=0.19\linewidth]{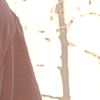}
		\label{fig:AblationStudy:HDRNet}
	}
	\subfloat[Ground truth]{
		\includegraphics[width=0.19\linewidth]{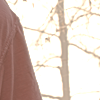}
		\label{fig:AblationStudy:Label}
	}
	\caption{Evaluation of different pipeline variants in an ablation study. Cascading separately trained Reconstruction- and Fusion-Subnets with optical flow based alignment (pre-align and post-align cascaded Merging-ISP) leads to accumulated ghosting artifacts related to misalignments. End-to-end learning both networks but replacing the alignment provided by Fusion-Subnet with optical flow (pre-align end-to-end Merging-ISP) leads to similar artifacts.}
	\label{fig:AblationStudy}
\end{figure}

Figure \ref{fig:AblationStudy:PreAlignCascading} and \ref{fig:AblationStudy:PostAlignCascading} depict both cascaded ISP variants in the presence of non-rigid motion between exposures. Both cascaded architectures suffer from error propagations in the form of color distortions. In the \textit{pre-align cascaded} approach, alignment is affected by missing pixel values resulting in ghosting artifacts. The \textit{post-align cascaded} approach is degraded by demosaicing errors affecting optical flow alignment and exposure merging. The end-to-end learned Merging-ISP shows higher robustness and reconstructs HDR data with less artifacts as depicted in Fig.~\ref{fig:AblationStudy:HDRNet}. Table~\ref{table:AblationStudy} confirms these observations quantitatively on the Kalantari test set. Overall, end-to-end learning leads to reconstructions with higher fidelity to the ground truth.

\subsubsection{Pre-aligning exposures vs. no alignment.}
The Fusion-Subnet in the proposed method is used to jointly align and merge a feature volume associated with multiple exposures. To further analyze the impact of this property, we evaluate an additional variant of Merging-ISP that employs optical flow alignment on its raw CFA inputs while using the same end-to-end learning for the Reconstruction-Subnet and the Fusion-Subnet (referred to as \textit{pre-align end-to-end} Merging-ISP). Figure~\ref{fig:AblationStudy:PreAlignHDRNet} depicts this pre-alignment approach. It is interesting to note that pre-alignment can cause errors that are difficult to compensate in the subsequent ISP stages. In contrast, handling exposure alignment in conjunction with low-level and high-level vision by the proposed method in Fig.~\ref{fig:AblationStudy:HDRNet} mitigates error accumulations. The benchmark in Table~\ref{table:AblationStudy} confirms that Merging-ISP without hand-crafted pre-alignment also leads to higher quantitative image quality.

\section{Conclusion}  
\label{sec:conclusion}

We proposed an effective deep neural network architecture named \textit{Merging-ISP} for multi-exposure HDR imaging of dynamic scenes. The proposed method outperforms the conventional approach of cascading different methods for image demosaicing, LDR to HDR conversion, exposure alignment, and merging both qualitatively and quantitatively. Our joint method avoids error propagations that typically appear in a cascaded camera ISP. The proposed Merging-ISP is also robust compared to state-of-the-art single-exposure HDR reconstruction in terms of obtaining HDR content. It avoids hand-crafted prepossessing steps like optical flow alignment of multiple exposures for merging in case of dynamic scenes.

In our future work, we aim at fine-tuning our architecture for additional ISP subtasks like denoising or sharpening. We also to plan to deploy and evaluate our method for HDR vision applications.

%
%
\bibliographystyle{splncs04}
\bibliography{main}

\end{document}